\documentclass{svmult}
\usepackage{epsfig,graphicx} % figures
\usepackage[bottom]{footmisc}% places footnotes at page bottom
\usepackage{natbib,url}      %RR additions
\bibpunct{(}{)}{;}{a}{}{,}   %RR reference punctuation
\def\cite#1{\citealp{#1}}    %RR restore old astroncite \cite command
\def\authorindex#1{}

\begin{document}\newcount\preprintheader\preprintheader=1

%%\setcounter{page}{1}
%RA to insert and reset to actual page number for your Astro-PH upload

%RR file: rr-assp-defs.tex = extra ASSP definitions by Rob Rutten
%RR last: Feb 24 2009 
%RR note: %RR Rob-to-Rob    
%RR site: cp rr-assp-defs.tex ~/rr/tex/macros/.

\def\thisvolume{these proceedings}
%RR OOPS: no volume number and no page numbers, Springer muckup

%RR journal abbreviations
%%%%%%%%%%%%%%%%%%%%%%%%%
\def\aj{{AJ}}			
\def\araa{{ARA\&A}}		
\def\apj{{ApJ}}			
\def\apjl{{ApJ}}		
\def\apjs{{ApJS}}		
\def\ao{{Appl.\ Optics}} 
\def\apss{{Ap\&SS}}		
\def\aap{{A\&A}}		
\def\aapr{{A\&A~Rev.}}		
\def\aaps{{A\&AS}}		
\def\an{{Astron.\ Nachrichten}}
\def\aspcs{{ASP Conf.\ Ser.}}
\def\assp{{Astrophys.\ \& Space Sci.\ Procs., Springer, Heidelberg}}
\def\azh{{AZh}}			
\def\baas{{BAAS}}		
\def\jrasc{{JRASC}}	
\def\memras{{MmRAS}}		
\def\mnras{{MNRAS}}
\def\nat{{Nat}}		
\def\pra{{Phys.\ Rev.\ A}} 
\def\prb{{Phys.\ Rev.\ B}}		
\def\prc{{Phys.\ Rev.\ C}}		
\def\prd{{Phys.\ Rev.\ D}}		
\def\prl{{Phys.\ Rev.\ Lett.}} %RR	
\def\pasp{{PASP}}
\def\pasj{{PASJ}}		
\def\qjras{{QJRAS}}
\def\science{{Sci}}		
\def\skytel{{S\&T}}		
\def\solphys{{Solar\ Phys.}} 
\def\sovast{{Soviet\ Ast.}}  
\def\ssr{{Space\ Sci.\ Rev.}}
\def\svassp{{Astrophys.\ Space Sci.\ Procs., Springer, Heidelberg}}
\def\zap{{ZAp}}			
\let\astap=\aap
\let\apjlett=\apjl
\let\apjsupp=\apjs
\def\grl{{Geophys.\ Res.\ Lett.}}  %RR Weiss
\def\jgr{{J. Geophys.\ Res.}} %RR Manoharan

%RR astronomy and math commands copied from ASP
%%%%%%%%%%%%%%%%%%%%%%%%%%%%%%%%%%%%%%%%%%%%%%%
\def\ion#1#2{{\rm #1}\,{\uppercase{#2}}}  %RR ~>\, \sc > uc 
\def\deg{\hbox{$^\circ$}}
\def\sun{\hbox{$\odot$}}
\def\earth{\hbox{$\oplus$}}
\def\la{\mathrel{\hbox{\rlap{\hbox{\lower4pt\hbox{$\sim$}}}\hbox{$<$}}}}
\def\ga{\mathrel{\hbox{\rlap{\hbox{\lower4pt\hbox{$\sim$}}}\hbox{$>$}}}}
\def\sq{\hbox{\rlap{$\sqcap$}$\sqcup$}}
\def\arcmin{\hbox{$^\prime$}}
\def\arcsec{\hbox{$^{\prime\prime}$}}
\def\fd{\hbox{$.\!\!^{\rm d}$}}
\def\fh{\hbox{$.\!\!^{\rm h}$}}
\def\fm{\hbox{$.\!\!^{\rm m}$}}
\def\fs{\hbox{$.\!\!^{\rm s}$}}
\def\fdg{\hbox{$.\!\!^\circ$}}
\def\farcm{\hbox{$.\mkern-4mu^\prime$}}
\def\farcs{\hbox{$.\!\!^{\prime\prime}$}}
\def\fp{\hbox{$.\!\!^{\scriptscriptstyle\rm p}$}}
\def\micron{\hbox{$\mu$m}}
\def\onehalf{\hbox{$\,^1\!/_2$}}	
\def\onethird{\hbox{$\,^1\!/_3$}}
\def\twothirds{\hbox{$\,^2\!/_3$}}
\def\onequarter{\hbox{$\,^1\!/_4$}}
\def\threequarters{\hbox{$\,^3\!/_4$}}
\def\ubv{\hbox{$U\!BV$}}		
\def\ubvr{\hbox{$U\!BV\!R$}}		
\def\ubvri{\hbox{$U\!BV\!RI$}}		
\def\ubvrij{\hbox{$U\!BV\!RI\!J$}}		
\def\ubvrijh{\hbox{$U\!BV\!RI\!J\!H$}}		
\def\ubvrijhk{\hbox{$U\!BV\!RI\!J\!H\!K$}}		
\def\ub{\hbox{$U\!-\!B$}}		
\def\bv{\hbox{$B\!-\!V$}}		
\def\vr{\hbox{$V\!-\!R$}}		
\def\ur{\hbox{$U\!-\!R$}}

%%%%%%%%%%%%%%%%%%%%%%%%%%%%%%%%%%%%%%%%%%%%%%%%%%%%%%%%%%%%%%%%%%%%%%%%%%%%
%RR RJR additional commands
%%%%%%%%%%%%%%%%%%%%%%%%%%%%%%%%%%%%%%%%%%%%%%%%%%%%%%%%%%%%%%%%%%%%%%%%%%%%

%RR -- non-bullet item marker in itemize list 
\def\labelitemi{{\bf --}}  

%RR -- latin abbreviations
\def\rmit#1{{\it #1}}              %% italics (RR style, Kluwer)
\def\rmit#1{{\rm #1}}              %% redefine for ASP, A&A, ApJ, Springer
\def\etal{\rmit{et al.}}           %% use \etal\ for space behind it        
\def\etc{\rmit{etc.}}           
\def\ie{\rmit{i.e.,}}              %% , required (Webster 1681)
\def\eg{\rmit{e.g.,}}              %% , required (Webster 1681)
\def\cf{cf.}                       %% no Latin, always Roman (Webster 1686)
\def\viz{\rmit{viz.}}
\def\vs{\rmit{vs.}}

%RR -- mathematical
\def\rot{\hbox{\rm rot}}
\def\div{\hbox{\rm div}}
\def\lesssim{\mathrel{\hbox{\rlap{\hbox{\lower4pt\hbox{$\sim$}}}\hbox{$<$}}}}
\def\gtrsim{\mathrel{\hbox{\rlap{\hbox{\lower4pt\hbox{$\sim$}}}\hbox{$>$}}}}
\def\dif{\: {\rm d}}                       %% differential d with space
\def\ep{\:{\rm e}^}                        %% e^ with space and roman e
\def\dash{\hbox{$\,-\,$}}                  %% math-like hyphen
\def\is{\!=\!}                             %% = in text for tighter spacing

%RR --stellar stuff
\def\starname#1#2{${#1}$\,{\rm {#2}}}  %% \starname{\alpha}{Cen~A} 
\def\Teff{\hbox{$T_{\rm eff}$}}   

%RR -- units (in addition to the ASP ones above)
\def\kms{\hbox{km$\;$s$^{-1}$}}
\def\ms{\hbox{m$\;$s$^{-1}$}}
\def\Mxcm{\hbox{Mx\,cm$^{-2}$}}    %% no 2, damn tex

%RR -- magnetic field 
\def\Bapp{\hbox{$B_{\rm app}$}}    %% apparent flux density, Lites convention

%RR -- oscillations
\def\komega{($k, \omega$)}                 %% k - omega 
\def\kf{($k_h,f$)}                         %% f - k_h
\def\VminI{\hbox{$V\!\!-\!\!I$}}           %% V-I
\def\IminI{\hbox{$I\!\!-\!\!I$}}           %% I-I
\def\VminV{\hbox{$V\!\!-\!\!V$}}           %% V-V
\def\Xt{\hbox{$X\!\!-\!t$}}                %% X-t

%RR -- atomic levels
%%      use:    \level 3s3p 3Pe
%%              \level 3s$^2$ {1,3}P{e,o}
%%              \level {} 3Ge
\def\level #1 #2#3#4{$#1 \: ^{#2} \mbox{#3} ^{#4}$}   

%RR -- some spectral species
\def\specchar#1{\uppercase{#1}}    %% to be redefined for A&A = \sc
\def\AlI{\mbox{Al\,\specchar{i}}}  %% use \AlI\ for space behind it
\def\BI{\mbox{B\,\specchar{i}}} 
\def\BII{\mbox{B\,\specchar{ii}}}  
\def\BaI{\mbox{Ba\,\specchar{i}}}  
\def\BaII{\mbox{Ba\,\specchar{ii}}} 
\def\CI{\mbox{C\,\specchar{i}}} 
\def\CII{\mbox{C\,\specchar{ii}}} 
\def\CIII{\mbox{C\,\specchar{iii}}} 
\def\CIV{\mbox{C\,\specchar{iv}}} 
\def\CaI{\mbox{Ca\,\specchar{i}}} 
\def\CaII{\mbox{Ca\,\specchar{ii}}} 
\def\CaIII{\mbox{Ca\,\specchar{iii}}} 
\def\CoI{\mbox{Co\,\specchar{i}}} 
\def\CrI{\mbox{Cr\,\specchar{i}}} 
\def\CriI{\mbox{Cr\,\specchar{ii}}} 
\def\CsI{\mbox{Cs\,\specchar{i}}} 
\def\CsII{\mbox{Cs\,\specchar{ii}}} 
\def\CuI{\mbox{Cu\,\specchar{i}}} 
\def\FeI{\mbox{Fe\,\specchar{i}}} 
\def\FeII{\mbox{Fe\,\specchar{ii}}} 
\def\FeIX{\mbox{Fe\,\specchar{ix}}}
\def\FeX{\mbox{Fe\,\specchar{x}}}
\def\FeXVI{\mbox{Fe\,\specchar{xvi}}}
\def\FrI{\mbox{Fr\,\specchar{i}}}
\def\HI{\mbox{H\,\specchar{i}}} 
\def\HII{\mbox{H\,\specchar{ii}}} 
\def\Hmin{\hbox{\rmH$^{^{_{\scriptstyle -}}}$}}      %% H^min, elegant
\def\Hemin{\hbox{{\rm He}$^{^{_{\scriptstyle -}}}$}} %% He^min, idem
\def\HeI{\mbox{He\,\specchar{i}}} 
\def\HeII{\mbox{He\,\specchar{ii}}} 
\def\HeIII{\mbox{He\,\specchar{iii}}} 
\def\KI{\mbox{K\,\specchar{i}}} 
\def\KII{\mbox{K\,\specchar{ii}}} 
\def\KIII{\mbox{K\,\specchar{iii}}} 
\def\LiI{\mbox{Li\,\specchar{i}}} 
\def\LiII{\mbox{Li\,\specchar{ii}}} 
\def\LiIII{\mbox{Li\,\specchar{iii}}} 
\def\MgI{\mbox{Mg\,\specchar{i}}} 
\def\MgII{\mbox{Mg\,\specchar{ii}}} 
\def\MgIII{\mbox{Mg\,\specchar{iii}}} 
\def\MnI{\mbox{Mn\,\specchar{i}}} 
\def\NI{\mbox{N\,\specchar{i}}}
\def\NIV{\mbox{N\,\specchar{iv}}}
\def\NaI{\mbox{Na\,\specchar{i}}}
\def\NaII{\mbox{Na\,\specchar{ii}}}
\def\NaIII{\mbox{Na\,\specchar{iii}}}
\def\NeVIII{\mbox{Ne\,\specchar{viii}}} 
\def\NiI{\mbox{Ni\,\specchar{i}}} 
\def\NiII{\mbox{Ni\,\specchar{ii}}}
\def\NiIII{\mbox{Ni\,\specchar{iii}}} 
\def\OI{\mbox{O\,\specchar{i}}} 
\def\OVI{\mbox{O\,\specchar{vi}}}
\def\RbI{\mbox{Rb\,\specchar{i}}} 
\def\SII{\mbox{S\,\specchar{ii}}} 
\def\SiI{\mbox{Si\,\specchar{i}}} 
\def\SiII{\mbox{Si\,\specchar{ii}}} 
\def\SrI{\mbox{Sr\,\specchar{i}}}
\def\SrII{\mbox{Sr\,\specchar{ii}}}
\def\TiI{\mbox{Ti\,\specchar{i}}} 
\def\TiII{\mbox{Ti\,\specchar{ii}}} 
\def\TiIII{\mbox{Ti\,\specchar{iii}}} 
\def\TiIV{\mbox{Ti\,\specchar{iv}}} 
\def\VI{\mbox{V\,\specchar{i}}} 
\def\HtwoO{\mbox{H$_2$O}}        %% H2O %RR TeX doesn't accept numbers alas
\def\Otwo{\mbox{O$_2$}}          %% O2

%RR -- hydrogen spectrum features
\def\Halpha{\mbox{H\hspace{0.1ex}$\alpha$}} %% \Halpha\ for space behind it
\def\Ha{\mbox{H\hspace{0.2ex}$\alpha$}}
\def\Hbeta{\mbox{H\hspace{0.2ex}$\beta$}}
\def\Hgamma{\mbox{H\hspace{0.2ex}$\gamma$}}
\def\Hdelta{\mbox{H\hspace{0.2ex}$\delta$}}
\def\Hepsilon{\mbox{H\hspace{0.2ex}$\epsilon$}}
\def\Hzeta{\mbox{H\hspace{0.2ex}$\zeta$}}
\def\Lyalpha{\mbox{Ly$\hspace{0.2ex}\alpha$}}
\def\Lybeta{\mbox{Ly$\hspace{0.2ex}\beta$}}
\def\Lygamma{\mbox{Ly$\hspace{0.2ex}\gamma$}}
\def\Lycont{\mbox{Ly\hspace{0.2ex}{\small cont}}}
\def\Baalpha{\mbox{Ba$\hspace{0.2ex}\alpha$}}
\def\Babeta{\mbox{Ba$\hspace{0.2ex}\beta$}}
\def\Bacont{\mbox{Ba\hspace{0.2ex}{\small cont}}}
\def\Paalpha{\mbox{Pa$\hspace{0.2ex}\alpha$}}
\def\Bralpha{\mbox{Br$\hspace{0.2ex}\alpha$}}

%RR -- Na D
\def\NaD{\mbox{Na\,\specchar{i}\,D}}    %% use \NaD\ for space behind it
\def\NaDone{\mbox{Na\,\specchar{i}\,\,D$_1$}}
\def\NaDtwo{\mbox{Na\,\specchar{i}\,\,D$_2$}}
\def\NaID{\mbox{Na\,\specchar{i}\,\,D}}
\def\NaIDone{\mbox{Na\,\specchar{i}\,\,D$_1$}}
\def\NaIDtwo{\mbox{Na\,\specchar{i}\,\,D$_2$}}
\def\Done{\mbox{D$_1$}}
\def\Dtwo{\mbox{D$_2$}}

%RR -- Mg b 
\def\Mgbone{\mbox{Mg\,\specchar{i}\,b$_1$}}
\def\Mgbtwo{\mbox{Mg\,\specchar{i}\,b$_2$}}
\def\Mgbthree{\mbox{Mg\,\specchar{i}\,b$_3$}}
\def\MgIb{\mbox{Mg\,\specchar{i}\,b}}
\def\MgIbone{\mbox{Mg\,\specchar{i}\,b$_1$}}
\def\MgIbtwo{\mbox{Mg\,\specchar{i}\,b$_2$}}
\def\MgIbthree{\mbox{Mg\,\specchar{i}\,b$_3$}}

%RR -- Ca II H & K 
\def\CaIIK{\mbox{Ca\,\specchar{ii}\,K}}       %% use \CaIIK\ for space
\def\CaIIH{\mbox{Ca\,\specchar{ii}\,H}}
\def\CaIIHK{\mbox{Ca\,\specchar{ii}\,H\,\&\,K}}
\def\HK{\mbox{H\,\&\,K}}
\def\Kthree{\mbox{K$_3$}}      %% numbers not permitted, alas
\def\Hthree{\mbox{H$_3$}}
\def\Ktwo{\mbox{K$_2$}}
\def\Htwo{\mbox{H$_2$}}
\def\Kone{\mbox{K$_1$}}     
\def\Hone{\mbox{H$_1$}}     
\def\KtwoV{\mbox{K$_{2V}$}}
\def\KtwoR{\mbox{K$_{2R}$}}
\def\KoneV{\mbox{K$_{1V}$}}
\def\KoneR{\mbox{K$_{1R}$}}
\def\HtwoV{\mbox{H$_{2V}$}}
\def\HtwoR{\mbox{H$_{2R}$}}
\def\HoneV{\mbox{H$_{1V}$}}
\def\HoneR{\mbox{H$_{1R}$}}

%RR -- Mg II h & k 
\def\hk{\mbox{h\,\&\,k}}
\def\kthree{\mbox{k$_3$}}    
\def\hthree{\mbox{h$_3$}}
\def\ktwo{\mbox{k$_2$}}
\def\htwo{\mbox{h$_2$}}
\def\kone{\mbox{k$_1$}}     
\def\hone{\mbox{h$_1$}}     
\def\ktwoV{\mbox{k$_{2V}$}}
\def\ktwoR{\mbox{k$_{2R}$}}
\def\koneV{\mbox{k$_{1V}$}}
\def\koneR{\mbox{k$_{1R}$}}
\def\htwoV{\mbox{h$_{2V}$}}
\def\htwoR{\mbox{h$_{2R}$}}
\def\honeV{\mbox{h$_{1V}$}}
\def\honeR{\mbox{h$_{1R}$}}

%%%%%%%%%%%%%%%%%%%%%%%%%%%%%%%%%%%%%%%%%%%%%%%%%%%%%%%%%% preprint header
%RR redefine @maketitle in svmult.cls to add slug on top
\ifnum\preprintheader=1     %RR ADAPT: 0 or 1 = preprintheader 
\makeatletter  %RR redefine symbol @ (trick from Pit Suetterlin)
\def\@maketitle{\newpage
\markboth{}{}%
%RR=================================
  {\em \footnotesize To appear in ``Magnetic Coupling between the Interior 
       and the Atmosphere of the Sun'', eds. S.~S.~Hasan and R.~J.~Rutten, 
       Astrophysics and Space Science Proceedings, Springer-Verlag, 
       Heidelberg, Berlin, 2009.}\par
%RR=================================
 \def\lastand{\ifnum\value{@inst}=2\relax
                 \unskip{} \andname\
              \else
                 \unskip \lastandname\
              \fi}%
 \def\and{\stepcounter{@auth}\relax
          \ifnum\value{@auth}=\value{@inst}%
             \lastand
          \else
             \unskip,
          \fi}%
  \raggedright
 {\Large \bfseries\boldmath
  \pretolerance=10000
  \let\\=\newline
% \@hangfrom{\@svsec}%
%%%  \@svsec
  \raggedright
  \hyphenpenalty \@M
  \interlinepenalty \@M
  \if@numart
     \chap@hangfrom{}%!!!
  \else
     \chap@hangfrom{\thechapter\thechapterend\hskip\betweenumberspace}%!!!
  \fi
  \ignorespaces
  \@title \par}\vskip .8cm
\if!\@subtitle!\else {\large \bfseries\boldmath
  \vskip -.65cm
  \pretolerance=10000
  \@subtitle \par}\vskip .8cm\fi
 \setbox0=\vbox{\setcounter{@auth}{1}\def\and{\stepcounter{@auth}}%
 \def\thanks##1{}\@author}%
 \global\value{@inst}=\value{@auth}%
 \global\value{auco}=\value{@auth}%
 \setcounter{@auth}{1}%
{\lineskip .5em
\noindent\ignorespaces
\@author\vskip.35cm}
 {\small\institutename\par}
 \ifdim\pagetotal>157\p@
     \vskip 11\p@
 \else
     \@tempdima=168\p@\advance\@tempdima by-\pagetotal
     \vskip\@tempdima
 \fi
}
\makeatother     %RR define @ back
\fi

\title*{Spectropolarimetry with CRISP at the Swedish 1-m Solar Telescope}

\author{A. Ortiz \and
L. H. M. Rouppe van der Voort}
\authorindex{Ortiz, A.}
\authorindex{Rouppe van der Voort, L. H. M.}

\institute{Institute of Theoretical Astrophysics, University of Oslo}

\maketitle

\setcounter{footnote}{0}  %RR Springer forgot this one (and much more)

%%%%%%%%%%%%%%%%%%%%%%%%%%%%%%%%%%%%%%%%%%%%%%%%%%%%%%%%%%%%%%%%%%%%%%

\begin{abstract} 

CRISP (Crisp Imaging Spectro-polarimeter), the new spectropolarimeter
at the Swedish 1-m Solar Telescope, opens a new perspective in solar
polarimetry.  With better spatial resolution (0.13\arcsec) than Hinode
in the Fe\,I~6302~\AA\ line and similar polarimetric sensitivity reached
through postprocessing, CRISP complements the SP spectropolarimeter
onboard Hinode.  We present some of the data which we obtained in our
June 2008 campaign and preliminary results from LTE inversions of a
pore containing umbral dots.

\end{abstract}

%%%%%%%%%%%%%%%%%%%%%%%%%%%%%%%%%%%%%%%%%%%%%%%%%%%%%%%%%%%%%%%%%%%%%%

%%%%%%%%%%%%%%%%%%%%%%%%%%%%%%%%%%%%%%%%%%%%%%%%%%%%%%%%%%%%%%%%%%%%%%
\section{Introduction}      \label{ortiz-introduction}

CRISP (CRisp Imaging Spectro-Polarimeter) is a new imaging
spectropolarimeter installed at the Swedish 1-m Solar Telescope (SST,
\cite{ortiz-2003SPIE.4853..341S}) in March 2008.
The instrument is based on a dual
Fabry-P\'erot interferometer system similar to that described by
\citet{ortiz-schar06}. It combines a high spectral resolution, high reflectivity etalon with
a low resolution, low reflectivity etalon. It has been designed as compact as possible, i.e., with a minimum of
optical surfaces, to avoid straylight as well as possible.

For polarimetric studies, nematic liquid crystals are used to modulate
the light. These crystals change state in less than 10~ms, which is
faster than the CCD read-out time. A polarizing beam splitter close to the focal plane splits the beam
onto two 1024$\times$1024 synchronized CCD's that measure the two orthogonal
polarization states simultaneously. This facilitates a significant
reduction of seeing crosstalk in the polarization maps. 

A third, synchronized, CCD camera records wide-band images through the
prefilter of the Fabry-P\'erot system. These images serve as an anchor
channel for Multi-Object Multi-Frame Blind Deconvolution (MOMFBD)
image restoration (\cite{ortiz-2005SoPh..228..191V}) which enables
near-perfect alignment between the sequentially recorded polarization
and line position images. For more details on MOMFBD processing of
polarization data see \citet{ortiz-2008A&A...489..429V}.

The etalons can sample spectral lines between 510 and 860~nm. The
field of view (FOV) is 70\arcsec$\times$70\arcsec; the pixel size
0.07\arcsec/pixel. The instrument has been designed to allow
diffraction-limited observation at 0.13\arcsec\ angular resolution in
the \FeI~630~nm lines.

\section{The June 2008 campaign data and processing}     \label{ortiz-data}

The data displayed here were recorded on June 12, 2008 as part of a
campaign during June 2008. The target was a pore (AR 10998) located at
S09\,E24 ($\mu$=0.79).  The field of view was 70\arcsec $\times$
70\arcsec.

\begin{figure} 
\centering
\includegraphics[width=\textwidth]{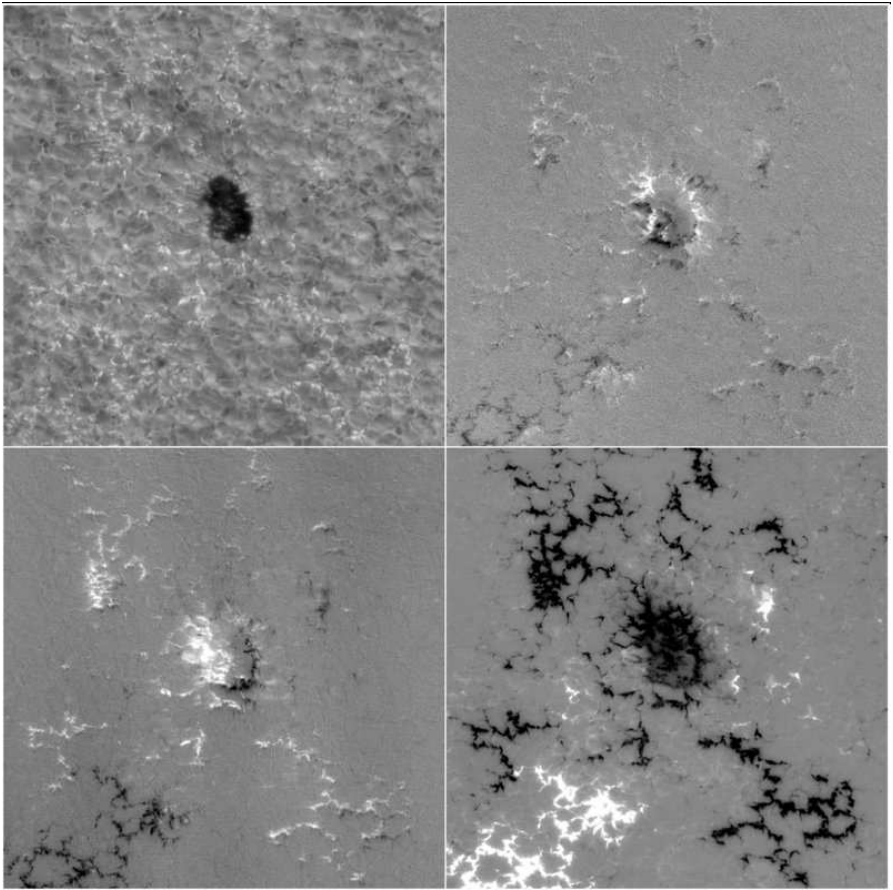}
\caption[]{\label{ortiz-fig1}
Clockwise: Stokes $I$, $Q$, $V$ and $U$
images taken in the blue wing of Fe\,I 6302.5~\AA\ at
$\Delta \lambda =-48$~m\AA, on June 12, 2008.}
\end{figure}

The images recorded correspond to complete Stokes measurements at 15
line positions in steps of 48 m\AA, from $-336$\,m\AA\ to +336\,m\AA,
in each of the \FeI\ lines, 6301.5 and 6302.5~\AA. In addition, images
were recorded at one continuum wavelength.  Each camera operated at
35~Hz frame rate. For each wavelength and LC state, 7 images were so
recorded per camera. Each sequence for subsequent MOMFBD processing
consists of about 870 images per CCD (2600 in total), recorded during
30~s.  The images were divided into overlapping 64$\times$64 pixel
subfields sampling different isoplanaic patches with overlaps.  All
images from each subfield were then processed as a single MOMFBD
set. They were demodulated with respect to the polarimeter and a
detailed telescope polarization model.  In addition, the resulting
Stokes images were corrected for remaining $I$ to $Q$, $U$ and $V$
crosstalk by subtraction of the Stokes continuum images.
Figure~\ref{ortiz-fig1} shows an example of the resulting Stokes
images.

\begin{figure} 
\centering
\includegraphics[width=7cm]{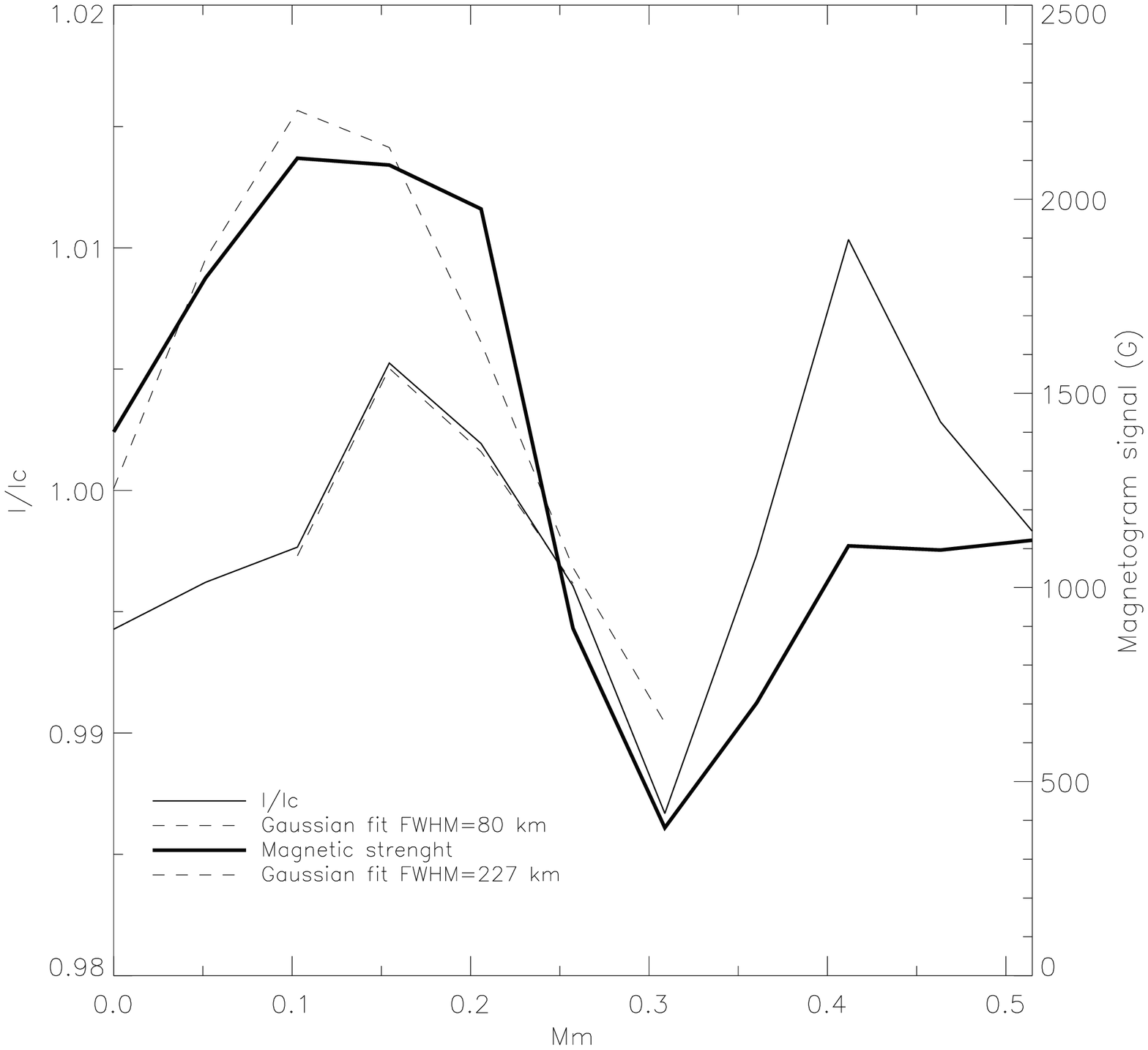}
\caption[]{\label{ortiz-fig2}
Cut along brightenings in the Stokes I image (thin line) and magnetic field obtained from inversions (thick line). We have fitted a gaussian to the smallest feature we can observe, both in the intensity image and the resulting magnetic field (dotted lines). The fits give us FWHM's of 80 km for I/Ic and 227 km for the magnetic field.}
\end{figure}

The theoretical diffraction limit of the SST is
$\lambda/D=0.13\arcsec$ at 6303~\AA.  We measured the real resolution
obtained in our June observations by identifying the smallest
intensity feature and fitting a Gaussian to it.
Figure~\ref{ortiz-fig2} shows a cut through a bright point with 80~km
FWHM for the Gaussian fit. This value is equivalent to 0.11\arcsec,
which is slightly lower than the theoretical resolution 0.13\arcsec\,
but consistent with it, due to the MOMFBD post-processing performed to
the data.  We estimated the noise level for the Stokes profiles to be
around $2\cdot10^{-3}$ for Stokes Q/I$_{c}$, U/I$_{c}$ and V/I$_{c}$.

\section{Inversions and results}    \label{ortiz-results}

To derive the atmospheric parameters from the observed Stokes images
we use a least-square inversion code, LILIA \citep{ortiz-hector01},
based on LTE atmospheres. We assume a one component, laterally
homogeneous atmosphere together with stray light contamination. The
inversions returns 9 free parameters as a function of optical depth,
including the three components of the magnetic field vector (strength,
inclination and azimut), LOS velocity and temperature among others. We
apply the inversion to both the \FeI\ 6301.5 and 6302.5~\AA\ lines
simultaneously.

\begin{figure} 
\centering
\includegraphics[width=\textwidth]{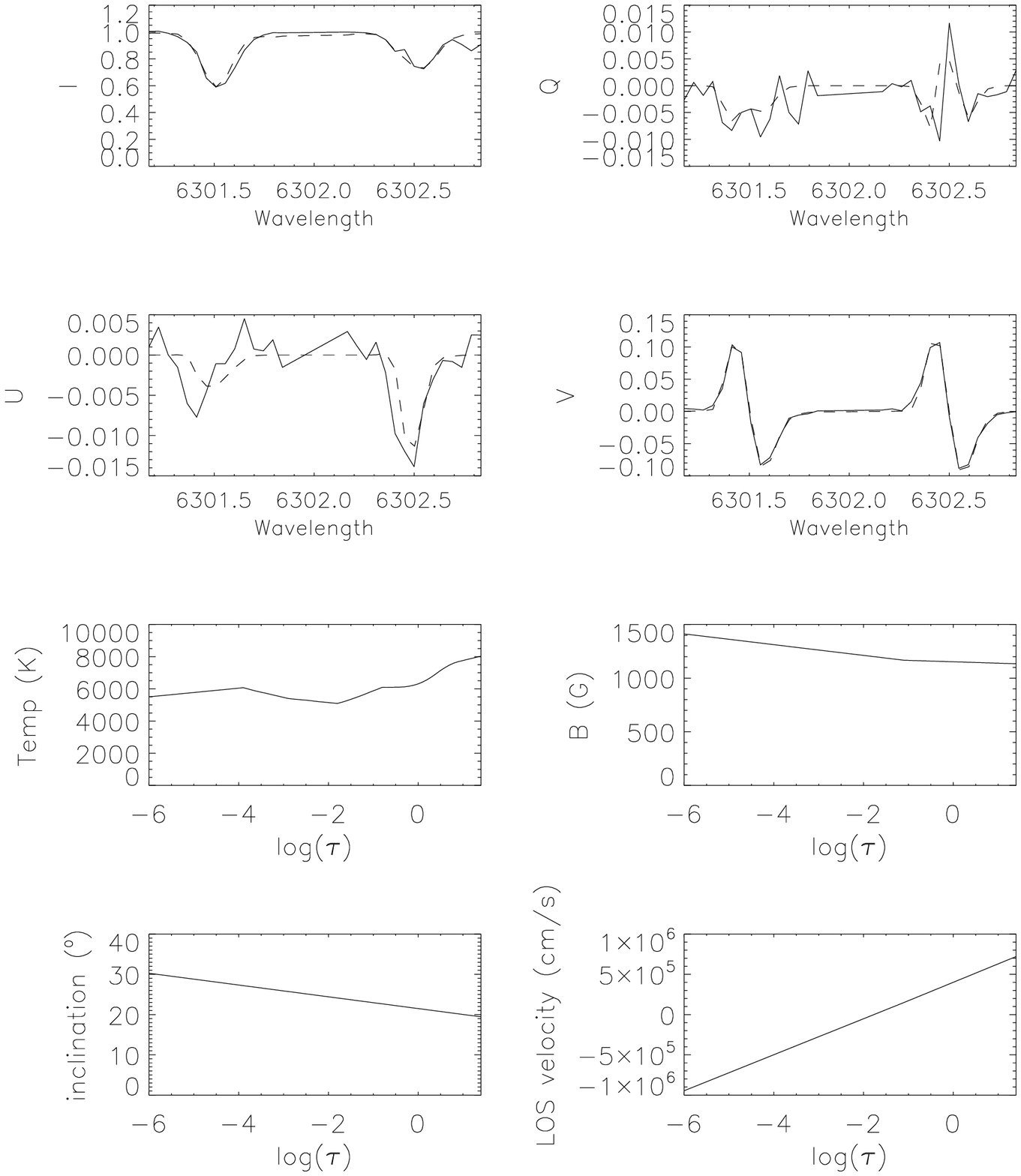}
\caption[]{\label{ortiz-fig3}
Results from the LILIA inversion of a bright point observed in an
intergranular lane. Observed (solid) and fitted (dashed) $I/I_c$,
$Q/I_c$, $U/I_c$ and $V/I_c$ profiles (upper panels), as well as
atmospheric parameters (temperature, magnetic field, inclination and
line-of-sight velocity) obtained through the inversion as a function
of optical depth (lower panels).}
\end{figure}

Figure~\ref{ortiz-fig3} shows an example of the inversion of an
individual pixel belonging to a bright point. In this particular case
the inversion code yielded a field strength of 1100 G, inclination of
25\deg, and LOS velocity of 0.6~\kms, (downflow) at $\log(\tau)=-1.5$.

\begin{figure} 
\centering
\includegraphics[width=\textwidth]{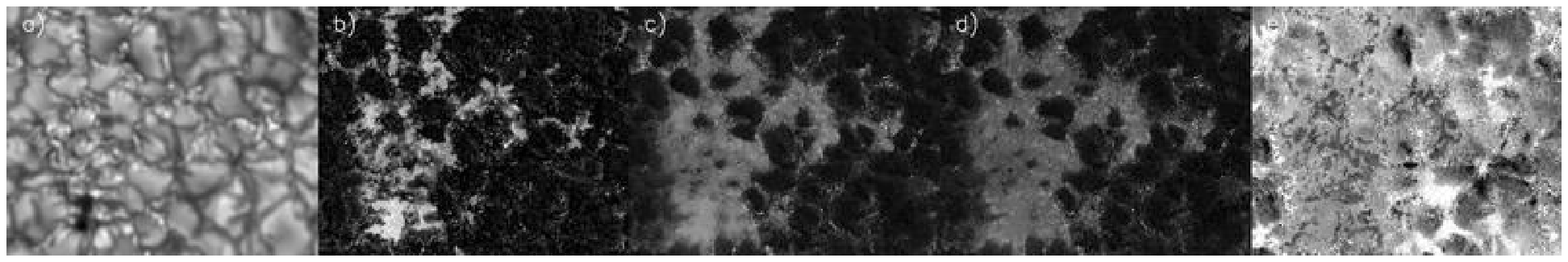}
\caption[]{\label{ortiz-fig4}
(a) Intensity image of a region containing a micro-pore and
intergranular magnetic fields including ribbons and bright points. (b)
(c) and (d): Mmagnetic field strength maps resulting from the inversion at
log($\tau$)=0, -1.5 and -2 respectively.  (e) LOS velocity map at
log($\tau$)=-2. Downflows are observed in the intergranular lanes,
while upflows are observed in the center of granules. The FOV is
14\arcsec $\times$ 10\arcsec.}
\end{figure}

\begin{figure} 
\centering
\includegraphics[width=\textwidth]{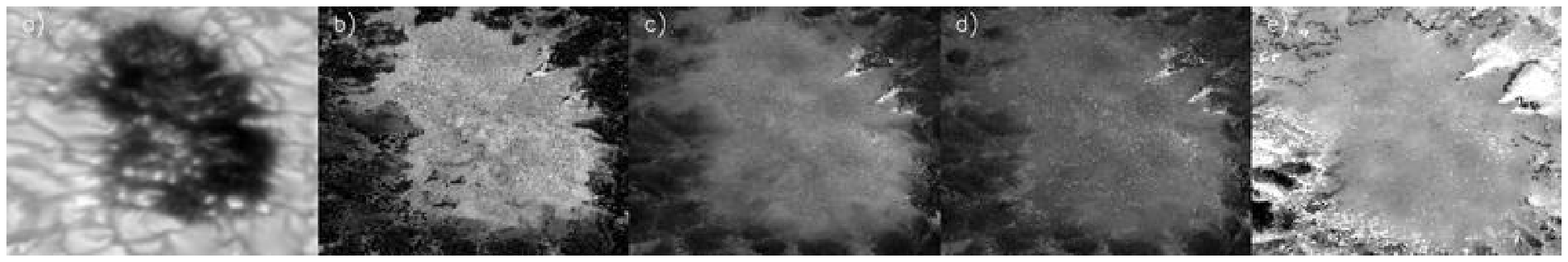}
\caption[]{\label{ortiz-fig5}
(a) Intensity image of a region containing a sunspot. Umbral dots and
other structures can be identified.  (b), (c) and (d):
magnetic field strength maps resulting from the inversion at
$\log(\tau)=0$, $-1.5$, and $-2$, respectively. (e) LOS velocity map at
$\log(\tau)=-2$. The FOV is 10\arcsec $\times$ 10\arcsec.}
\end{figure}

Figures~\ref{ortiz-fig4} and \ref{ortiz-fig5} show maps of the
obtained magnetic field strength and line-of-sight (LOS) velocity at
different heights. Figure~\ref{ortiz-fig4} shows a micro-pore as well
as brightenings produced by emergent magnetic fields. Ribbons
(\cite{ortiz-2004A&A...428..613B}) can be distinguished. Upflows are
correlated with the positions of the center of the granules, while
downflows are correlated with the intergranular lanes, except in those
areas where the magnetic field is emerging, in which velocities are
lower due to the supression of convection. Figure~\ref{ortiz-fig5}
presents a pore with several umbral dots and structures within. These
brighter umbral structures show lower magnetic field strengths than
the darker parts of the umbra as well as higher
temperatures. Velocities inside the pore tend to be lower than in
normal granulation patterns. The velocity field of the umbral
structures will be analyzed in a future publication.

\section{Discussion}                   \label{ortiz-conclusions}

We have presented the first LTE inversions based on data obtained with the CRISP imaging spectropolarimeter, used with the 1-m SST. The spatial resolution of the Stokes data presented here represents a major improvement compared to other ground-based data and even to the recent Hinode data. We have shown here both the capabilities of the CRISP instrument and of the inversion code applied to this data. We expect to make use of such capabilities for exploring in detail the umbral dots and other structures found in the micro-pore present on June 12, 2008. 

\begin{acknowledgement}

The Swedish 1-m Solar Telescope is operated on the island of La Palma by the Institute for
Solar Physics of the Royal Swedish Academy of Sciences in the Observatorio del Roque de
los Muchachos of the Instituto de Astrof\'{\i}sica de Canarias.

\end{acknowledgement}

%%%%%%%%%%%%%%%%%%%%%%%%%%%%%%%%%%%%%%%%%%%%%%%%%%%%%%%%%%%%%%%%%%%%%%%%%%% References
%%%%%%%%%%%%%%%%%%%%%%%%%%%%%%%%%%%%%%%%%%%%%%%%%%%%%%%%%%%%%%%%%%%%%%%%%

\begin{small}

\bibliographystyle{rr-assp}       %RR hacked from aa.bst
\bibliography{/tmp/adsfiles.bib,bangalore.bib}

\end{small}

\end{document}